\documentclass[a4paper,11pt]{article}
\usepackage{authblk}
\usepackage[utf8]{inputenc}
\usepackage[scale=0.8]{geometry}
\usepackage{setspace}
\usepackage{graphicx}

\date{}
\title{Continuous Wire Reinforcement for Jammed Granular Architecture}

\author{Matthias Fauconneau}
\author{Falk K. Wittel}
\author{Hans J. Herrmann}
\affil{Computational Physics for Engineering Materials, ETH Z{\"u}rich, Stefano-Franscini-Platz 3, 8093 Z{\"u}rich, Switzerland}

\begin{document}
\maketitle

\begin{abstract}
The mechanical behavior of continuous fiber reinforced granular columns is simulated by means of a Discrete Element Model. Spherical particles are randomly deposited simultaneously with a wire, that is deployed following different patterns inside of a flexible cylinder for triaxial compression testing. We quantify the effect of three different fiber deployment patterns on the failure envelope, represented by Mohr-Coulomb cones, and derive suggestions for improved deployment strategies. 
\end{abstract}
\section{Introduction}\label{sec:intro}
\begin{figure}[htb]
\centering{
  \includegraphics[width=8cm]{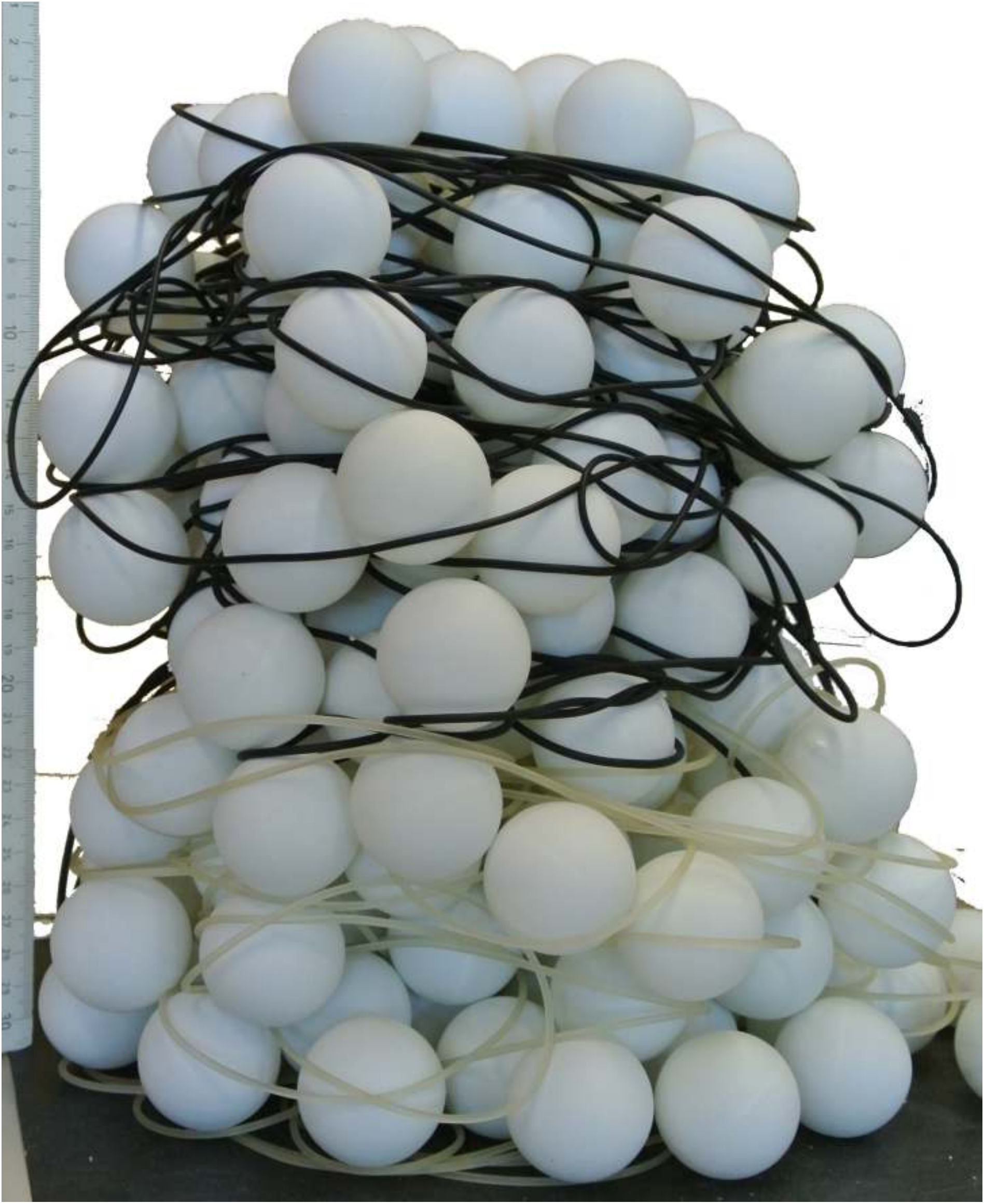}
\caption{Free standing column of table tennis balls (diameter 40~mm) with continuous reinforcement of wires (diameter 2~mm).\label{fig:pingpong}}}
\end{figure}

Granular systems of sand or gravel can either yield and flow, or be jammed and rigid. Materials, whose mechanical integrity results from a jamming process belong to a class of material that has been called fragile matter \cite{CATES_ETAL_98}. They show solid behavior when particles are driven into a jammed state by an external load, thus trapping or frustrating the disordered system. In rockfill structures for example, such a jammed state is desired. The fundamental difference and also challenge for architecture out of fragile matter with respect to cohesive materials is that upon changes of external load or when vibrated, it might unjam and the solid might apparently melt. In principle, different possibilities are commonly implemented to avoid this:
\begin{itemize}
\item By external confinement, e.g. in stacked gabion cages filled with gravel or encased granular columns with geogrids for improving the bearing capacity of weak soft soils \cite{N-K-YAH-11}.
\item Introducing attractive interactions by cementing particles.
\item By internal confinement with tensile elements like \mbox{two-dimensional} sheets of geogrids to obtain laminated reinforced granular columns \cite{WU-HONG-08,H-R-ALMEIDA-14}. 
\item By tensile one-dimensional objects, such as fibers, membrane strips, or hooked elements that provide the lateral tensile forces needed for preventing the buckling of the force chains \cite{Nozoe-etal-2013}. Such elements can be in-between particle contacts or attached to a certain number of particles.
\item By interlocking particles with concave surfaces. This way the volumetric strain becomes more dilatant upon compression, and larger tensile stresses can be propagated. Systems of this type were recently studied in Ref.~\cite{Jaeger-2014}.
\item By grading particle size and shape, for example radially for a granular column.
\item By combinations of the above.
\end{itemize}
For sand, the effect of mixing synthetic fibers, so-called Georobes, on the mechanical performance and failure behavior was studied extensively in the past \cite{NOZOE-ETAL-13,GUERRERO-VALLEJO-10,MICHALOWSKI-ZHAO-96}, resulting in patented products such as Texsol$\textsuperscript{\textregistered}$, where sand is mixed with continuous polyester resin fibers with fiber weight content varying between \mbox{0.1-0.2\%}. The fiber-reinforcement typically increases the cohesion without significantly changing the angle of internal friction \cite{GUANGXIN-ZHAO-03}. Hence tensile strength and the respective ultimate strain increase, resulting in gradual, more ductile failure behavior. Up-scaling from sand to rockfill material for architectural purposes is rather demanding. Randomly oriented fibers are replaced by layers of wires whose role inside the granular packing is very different from the one in sand. For confined fiber-reinforced sand, the bridging of shear bands by fibers is the most important effect, while for free standing granular columns made of large granulates, the hindering of lateral force chain buckling is the dominating mechanism. In a first study \cite{Petrus-2015}, various types of internal reinforcement with chains, wires, cable straps and many others were studied with respect to their applicability in robotic fabrication and overall performance. Wire reinforcement proved to be the most flexible and at the same time efficient way of reinforcing granulates. This principle applied to columns is the focus of our numerical study. A small demonstration is shown in Fig.~\ref{fig:pingpong}, where a free standing column of table tennis balls is obtained by reinforcing with a rubber wire.

Forces in granular matter are in general transmitted only through individual particle contacts and form force networks that are the key to jamming and yielding of granular assemblies. From the analysis of networks it becomes evident, that a large number of particles does not contribute to the mechanical force network~\cite{corwin-jaeger-nagel-2005,Majmudar-Behringer-1998}. We call them spectator particles as opposed to force network particles. In principle, spectator particles could be removed without observable consequences. However in a granular packing, already small changes can lead to significant rearrangements in the network topology, suddenly causing a spectator to become an important force network particle. Therefore, spectator particles give robustness to a jammed packing, as they avoid buckling of force chains by laterally supporting rows of force network particles~\cite{CATES_ETAL_98}. The final aim is to derive design principles for structural elements, such as columns and walls out of fragile matter that are robust and remain in a jammed state throughout their life by the dead load of the structure itself. This challenge can only be met after understanding the principles of granular interactions along with features of the force networks. Tensile reinforcements e.g. by continuous wires are important for obtaining robust structures. However, the effect of continuous wire reinforcement is twofold: On one hand, it is the element that allows for jamming of free standing columns, as it avoids lateral force chain buckling at the surface. On the other hand, wires can suppress force network reorganizations inside the bulk. 

To study the effect of continuous wire reinforcement we apply Discrete Element Methods (DEM) for models composed of spherical particles, an elastic wire and an elastic membrane confinement for pressurization in triaxial compression testing. After describing the model, we introduce three different wire deployment strategies and quantify their effect on the macroscopic systems behavior by estimating the respective Mohr-Coulomb limit surfaces.
\section{DEM compression simulations with wires}\label{sec:MuM}
\begin{figure}[htb]
\centering{
\includegraphics[width=8cm]{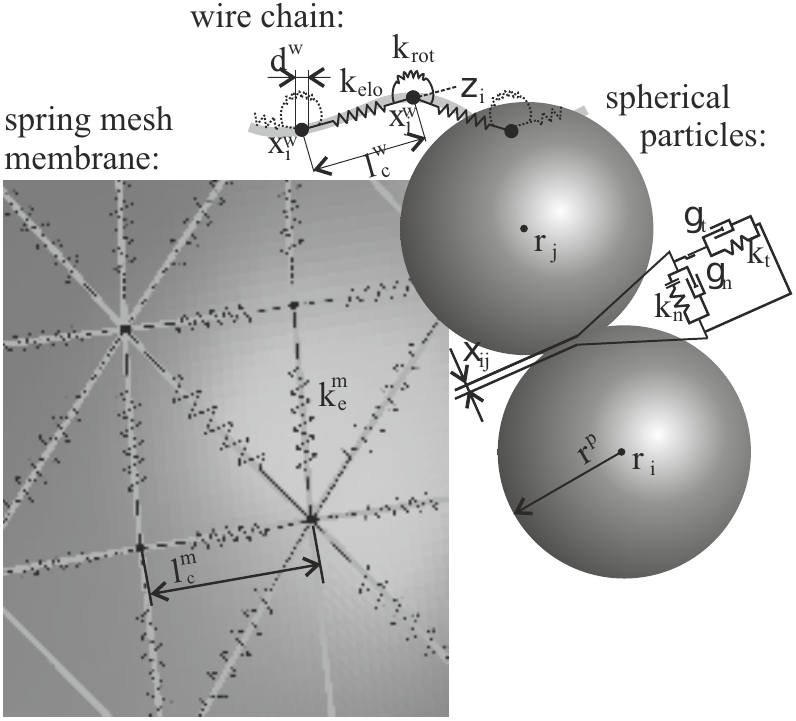}
\caption{Representation of each component of our model (membrane, wire and particles) using systems of connected masses, springs and dampers.\label{fig:syselements} } }
\end{figure}

Discrete Element Models have been successfully applied since they were first introduced by Cundall and Strack \cite{CUNDALL1979} to study rock mechanics. Applications of this soft-particle method range from static structures, to impact and explosive loading, using elementary particles of various shapes \cite{LUDING-08,herrmann1998modeling}. Typical scenarios involve shear failures under confining pressure with shear band formation, large disorder and rolling of particles. Newton's equations of motion govern the translational and rotational motion of the elements driven by forces and torques that arise from gravity, external pressure and contact interactions. Our three-dimensional (3D) model incorporates three different mechanical elements, namely, spherical particles, the continuous wire and the elastic membrane confinement (see Fig.~\ref{fig:syselements}), that are described in the following.

All \textbf{spherical particles} have identical radius $r^p$, Young's modulus $E^p$, Poisson's ratio $\nu^p$ and density $\rho^p$. The repulsive force between spheres due to the elastic contact is calculated by the Hertz law as a function of the overlapping distance $\xi$. With the vector $\vec{r}_{ij}$, connecting the centers of spheres $i$ and $j$, $\xi_{ij} = 2r^p - \left|\vec{r}_{ij}\right|$, the contact force is given by
\begin{equation}\label{eq:1}
\vec{F}_{ij}^{n} = \frac{4}{3}\frac{E^p\sqrt{R_{eff}}}{(1-(\nu^p)^2)}(\vec{\hat{r}_{ij}}\xi_{ij}^{3/2} - \gamma_n\sqrt{\xi_{ij}}\frac{d\xi_{ij}}{dt} ).
\end{equation}
Different sphere radii and types are considered via the effective radius $R_{eff}^{-1} = 1/R_i + 1/R_j$ and the effective Young's modulus  $E^{-1} = 1/E_i + 1/E_j$. Damping reduces $F_{ij}^n$ by the normal damping coefficient $\gamma_n$ and the deformation rate $d\xi_{ij}/dt$ of the contact. Eq.~\ref{eq:1} is used for all interactions where spheres are involved like grain-grain, wire-grain, wire-wire, membrane-grain and contact with the horizontal walls with $R_j=\infty$. Tangential contact is implemented via Coulomb's law. A spring with contact spring stiffness $k_t$ is attached between contact points of contacting spheres, exerting a restoring force for tangential displacements to mimic static friction that needs to be overcome for slip. If the spring force exceeds the dynamic friction force $F_t = \mu F_n $ with friction coefficient $\mu$ or if its elongation exceeds a threshold value $d_{cr}$, the contact spring is removed and particles are allowed to slip with a shear damping $\gamma_t v_{rel}^t$ that grows linearly with the relative tangential velocity $v_{rel}^t$ at the contact point. Note that displacements and rotations of involved particles contribute to $v_{rel}^t$. A detailed description of the contact model can be found in Ref.\~cite{LUDING-08}.

The \textbf{wire} with circular cross section of radius $r^w$ consists of $N^w$ wire segments represented by longitudinal springs with identical segment length $l^w$ and resulting stiffness $k_{elo}=E^w\pi(r^w)^2/l^w$. Bending resistance is introduced via rotational springs of stiffness $k_{rot}=E^w\pi (r^w)^4/(4l^w)$ attached to each node. The force from the longitudinal springs between the wire nodes $i$ and $j$ is
\begin{equation}\label{eq:tension}
\vec{F}_{ij}^{tension} = -k_{elo}\varepsilon\vec{\hat{r}_{ij}},
\end{equation}
where $\varepsilon = \left(\left|\vec{r}_{ij}\right| - l^w\right)/l^w$, with the unit vector $\vec{\hat{r}}_{ij}$ pointing from node $i$ to $j$. The bending moment at node $i$ reads $M_i=k_{rot}\zeta_i$ with the bending angle $\zeta_i$ given by the two segments attached to node $i$ (see Fig.~\ref{fig:syselements}). It is introduced via the respective force couples with the neighboring nodes. Self-contact of the wire, as well as the wire contact with particles is avoided by fictitious contact spheres with radius $r^w$ located at each wire node, treated by Eq.~\ref{eq:1}. Note that presently only vertex-sphere interactions are considered that require significantly smaller segment lengths than the spherical particle radius to achieve accurate contact forces. Through the implementation of spherocylinders, as well as beam elements \cite{spherocylinder}, the accuracy could be increased in the future.

Triaxial compression tests on granular systems require stress-controlled boundaries \cite{ASTM_D2850} that are realized in experiments by a rubbery \textbf{membrane} that seals the sample from the outside and thus allows for pressurization. In past DEM studies, the flexible membrane was modeled by chains of discrete elements \cite{Wang-Tonon-2009,Lee-etal-2012}, projecting membrane spheres on a cylinder \cite{Cui-Sullivan-ONeill-2007,Sullivan-Cui-ONeill-2008}, or by identifying the contour on which to apply a constant pressure \cite{PhysRevE.66.021301}. We represent the elastic, cylindrical membrane of thickness $d^m$ around the sample by a fine, regular triangular mesh with initial mesh size $l^m_c$ and point masses $m^m = \sqrt{3}/2 (l^m_c)^2 \rho^m d^m $ located at the nodes, using the density of the membrane $\rho^m$. Nodes are connected by springs of stiffness $k^m= \sqrt{3}/2 E^m d^m$ with the elastic modulus $E^m$. Contact with the particles is implemented in the same way as described for the wire with fictitious contact spheres of radius $d^m/2<r^m$ located at each membrane mesh node. Self-contact and contact with wire segments not considered so far. The upper and lower rim of the cylindrical membrane are fixed throughout the simulation. The external pressure is applied normal to the triangular faces and the respective forces are equally shared between the triangle nodes.

The time evolution of the system is followed by simultaneously integrating the equations of motion for the translation and rotation of all mass points assigned to spheres, wire, and membrane, using a $5^{th}$-order Gear predictor-corrector algorithm with time increment $dt$, as the upper wall moves towards the lower one at a constant displacement rate. Note that no viscous damping is implemented. For the angles, we use a quaternion representation \cite{Rapaport2004,Poschel2005}. All model parameters are summarized in Table~\ref{tab:properties}.
\begin{table}[htb]
\centering{
\caption{\label{tab:properties}Summary of model parameters.}
\begin{tabular}{llll}
\textbf{Particles:}\\
stiffness& $E^p$ & 10 & GPa\\
Poisson's ratio& $\nu^p$ & 0.3 & - \\
radius & $r^p$ & 1.23 & mm\\
density & $\rho^p$ & 7800 & kg/m$^3$\\ 
\textbf{Wire:}\\
stiffness& $E^w$ & 1 & GPa\\
Poisson's ratio& $\nu^w$ & 0.0 & - \\
segment length & $l_c^w$ & 1.25 & mm\\
radius & $r^w$ & 0.5 & mm\\
density & $\rho^w$ & 1000 & kg/m$^3$\\ 
\textbf{Membrane:}\\
stiffness& $E^m$ & 0.1 & GPa\\
Poisson's ratio & $\nu^m$ & 0.0 & - \\
mesh size & $l^m_c$ & 1.25 & mm\\
thickness & $d$ & 1 & mm\\
density & $\rho^m$ & 1000 & kg/m$^3$\\ 
\textbf{Interaction:}\\
friction coefficient & $\mu$ & 0.3 & -\\
length threshold & $d_{cr}$ & 0.1 & mm\\ 
normal damping coef. & $\gamma_n$ & 0.02 & s$^-1$\\ 
tang. damping coeff. & $\gamma_t$ & 0.015 &s$^-1$\\
contact spring stiffness & $k_t$ & $10^5$ & m$^{-1}$\\ 
boundary plate stiffness & $E^b$ & 1 & GPa\\
\textbf{System:}\\
number particles & $N^p$ & 1000 & -\\ 
number wire segments & $N^w$ & 8000 & -\\
number mesh nodes & $N^s$ & 20000 & -\\ 
initial cylinder radius & $R$ & 2 & cm\\
gravitational constant & $g$ & 10 & m/s$^2$\\
compression rate & $v_z$ & 0.2 & mm/s\\
time increment & $\delta t$ & 10 & $\mu$s\\
\end{tabular}}
\end{table}

\begin{figure}[htb]
\centering{
  \includegraphics[width=8cm]{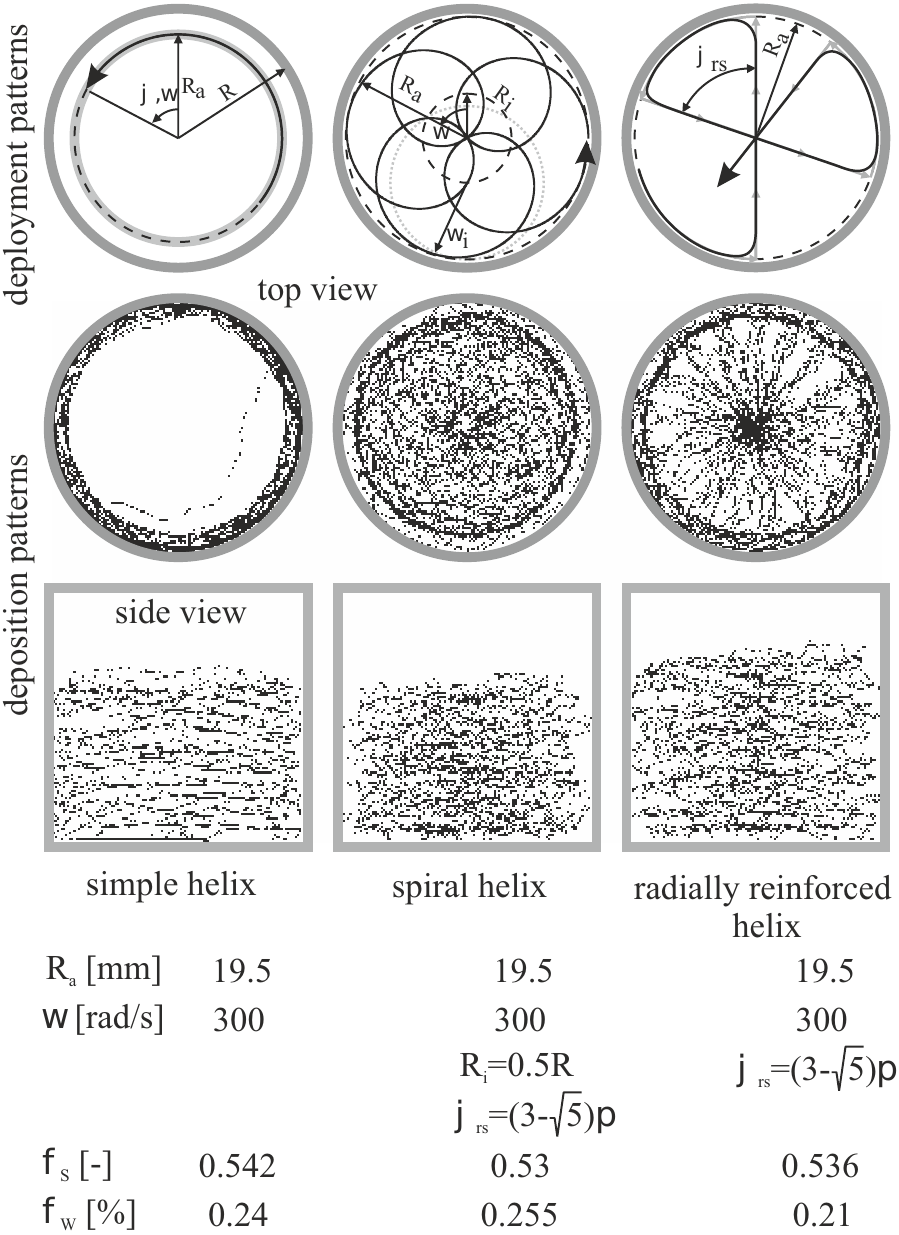}
\caption{The final wire pattern, in top and side view, respective deployment parameters and resulting packing fractions $\phi_S$ of spheres and volume fractions of wire $\phi_w$ for three different wire deployment patterns: simple helix, spiral helix, and radially reinforced helix. For all patterns the vertical deployment speed is set to $v_z^w=0.5$~m/s and the angular velocity on the helix radius to $\omega=300$~rad/s.\label{fig:deployment_pattern} }}
\end{figure}

The system preparation closely follows the rock printing procedure described in Ref.~\cite{Petrus-2015}. Particles are randomly deposited simultaneous with wire that is deployed along predefined trajectories. Note that gravitational forces $\vec{F}^{vol}_i=V_i\rho^{w/p} \vec{g}$, calculated with gravitational acceleration $\vec{g}$ act on both, wire and spheres with respective volume $V$ and density $\rho^{w/p}$, but not on the membrane. We study the effect of three different deployment strategies (see Fig.~\ref{fig:deployment_pattern}) namely (a) simple helix, (b) spiral helix with the center of the helix being located on a helix itself, and (c) a helix with periodic radial reinforcement. They are parametrized as follows: case (a) by vertical speed $v_z^w$, angular speed $\omega$ and helix radius $R_a$; case (b) by $v_z^w$, $R_a$ and radius of the inner helix radius $R_i$, $\omega$ and the angle for a rotation on a inner helix $\phi_{rs}$ from which the angular speed of the inner helix $\omega_i=2\pi/\phi_{rs}$ is calculated; and case (c) by $R_a$, $\omega$ and the radial reinforcement angle $\phi_{rs}$. The angular velocity $\omega$ refers to the deployment of the simple helix and its respective deployment speed is calculated on the helix radius $R_a$. It is kept constant for all deployment patterns. Note that $v_z^w$ and $\omega$ determine the volume fraction of the wire $\phi_w$. During the system preparation, the membrane is ideally rigid as membrane node positions are not updated. Depending on the deployment pattern and parameter choice, different packing ratios and fiber volume contents are achieved. We notice that any sharp corners in the deployment trajectories of the radially reinforced configuration (see Fig.~\ref{fig:deployment_pattern}) are not present after deposition due to the implementation of wire bending. However, for case (c) significant differences in curvature remain.
\section{Simulation of triaxial compression}
The compressive strengths are determined from triaxial compression tests at different confining pressures $\sigma_3$. The upper plate is vertically displaced at a constant rate of $v_z$. For a correct evaluation of the differential stress $\sigma_1-\sigma_3$, the pressure due to stretching of the elastic membrane around the granular packing has to be added. For this purpose we do not use $\sigma_3$, but the effective pressure as the sum over all forces of grain-membrane contacts normalized by the mean membrane area. The area is calculated from the distance between horizontal plates and the average radial distance of contacting particles from the vertical axis. We measure the force acting on both horizontal plates, normalized by the true area of the column, calculated from the average radial positions of all particles contacting the membrane. The maximum force measured during the triaxial test defines the yield strength. Values obtained for different confining pressures $\sigma_3$ are required to construct limit surfaces from Mohr's circles, i.e with diameters corresponding to the deviatoric stress. A linear regression through all maximum shear stress points gives the internal friction angle as slope and the cohesion as intersection with the shear stress axis.
\begin{figure}[htb]
\centering{
\includegraphics[width=8cm]{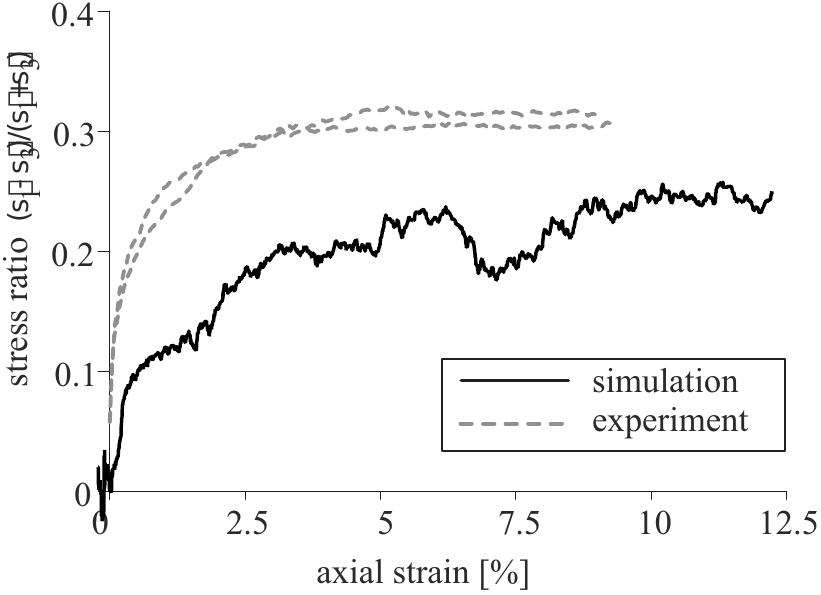}
\caption{\label{fig:validation}Comparison of two experiments from O'Neill~\cite{ONeill-2005} with simulation results without wire reinforcement.} }
\end{figure}
\subsection{Validation of the triaxial test}
\begin{figure}[htb]
\centering{
\includegraphics[width=8cm]{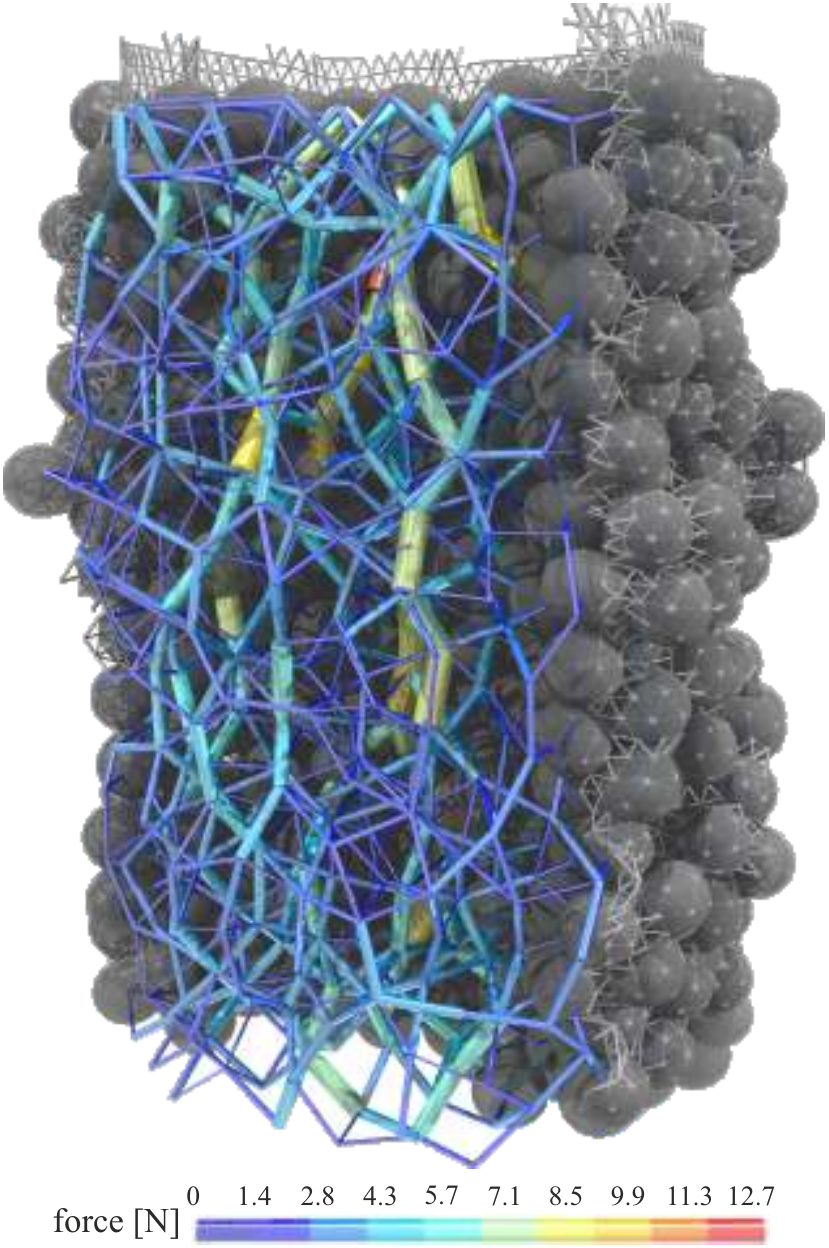}
\caption{\label{fig:forcenetwork} Resulting force network for the unreinforced case.}}
\end{figure}

We compare our simulations without wire to experimental results of O'Neill~\cite{ONeill-2005}, who performed triaxial compression tests on particle packings with equally sized steel spheres. One observes a similar profile in normalized stress ratio $(\sigma_1-\sigma_3)/(\sigma_1+\sigma_3)$ during axial compression, see Fig.~\ref{fig:validation}.  With increasing displacement, the forces on the static friction contacts increase and trigger slip events, resulting in particle reconfigurations and a load drops. Eventually, an asymptotic stress ratio is reached. The force network, visualized in Fig.~\ref{fig:forcenetwork} shows that the dominant force lines are close to the center of the sample. Note that differences arise mainly due to the smaller particle number with respect to the experiments.
\subsection{Triaxial tests with different wire patterns}
\begin{figure*}[htb]
\centering{
\centering{\includegraphics[width=16cm]{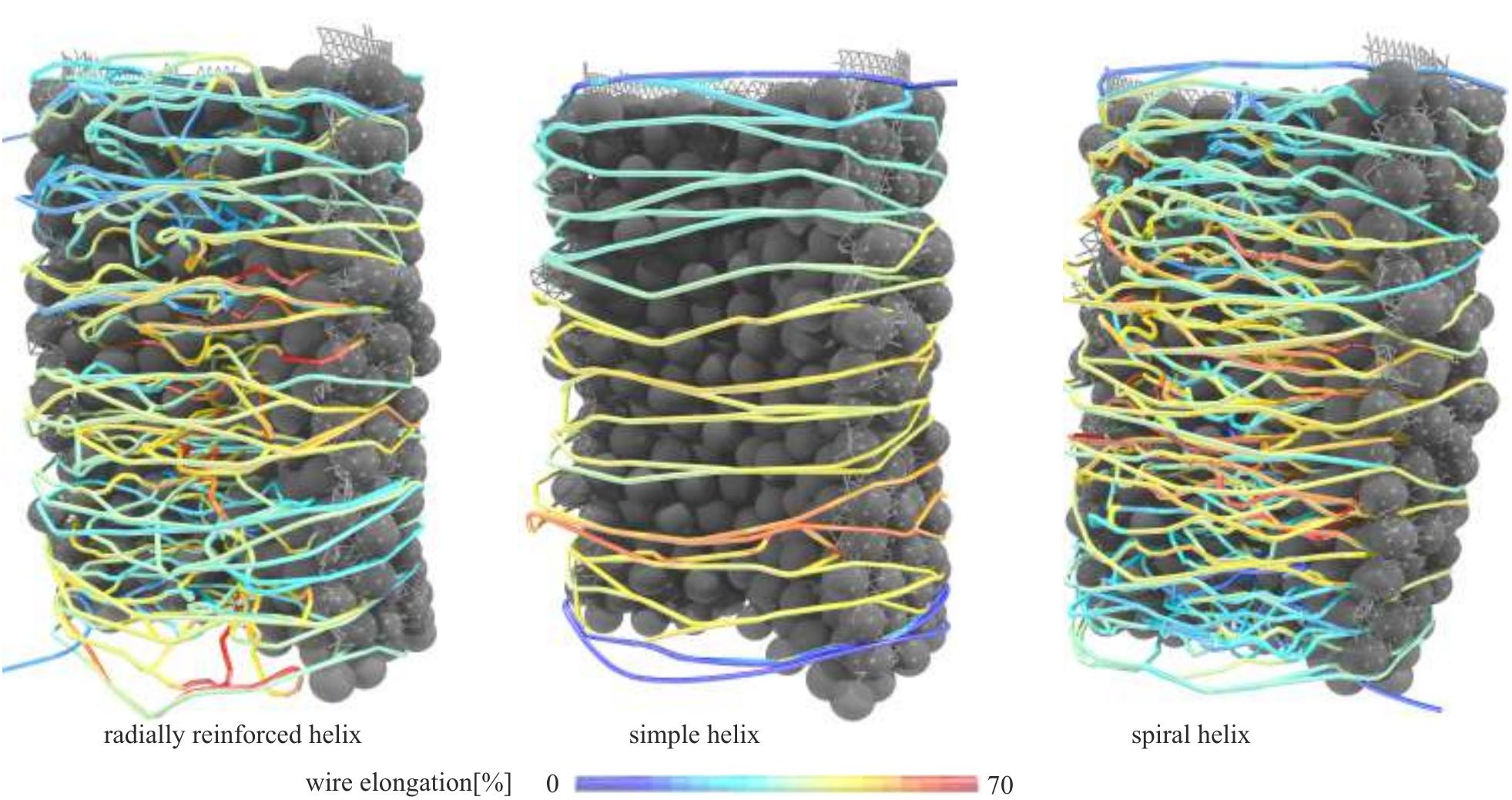}}
\caption{\label{fig:all}Snapshots of configurations at 10\% axial strain with identical color legend.} }
\end{figure*}

In order to evaluate the effect of different wire patterns on macroscopic material performance, we ran the triaxial tests on the three selected representative patterns, described in Sec.~\ref{sec:MuM}. Fig.~\ref{fig:all} shows the samples with 10\% axial strain for a confining pressure of $\sigma_3=80$~KPa in a soft membrane cylinder of radius $R=0.02$~m. Colors represent the elongation of wire segments. For visualization purposes, in a wedge shaped region, only wire is shown. Table.~\ref{tab:fraction} gives the grain and wire volume fractions. Slight differences can be noted due to the way the wire deposition affects the granular packing.
\begin{table*}[htb]
\centering{
\caption{\label{tab:fraction}Volume of sample and volume fractions of grain and wire for the different patterns.}
\begin{tabular}{llll}
             & Volume [$cm^3$] & Grain fraction [\%] & Wire fraction [\%] \\ 
Without wire & 146 & 44  & 0  \\
Simple helix pattern & 150 & 43 & 4.9\\
Spiral helix pattern & 152 & 43 & 5.1\\
Radially reinforced helix pattern & 158 & 41 & 4.9 \\
\end{tabular}}
\end{table*}

\begin{figure}[htb]
\centering{
\includegraphics[width=8cm]{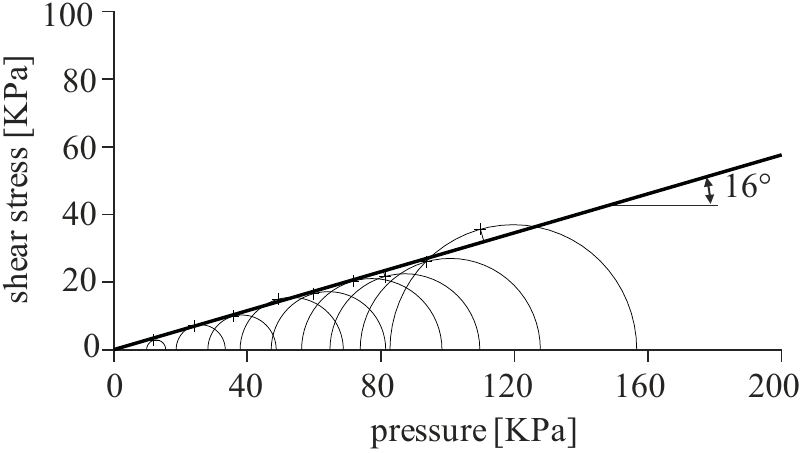}
\caption{\label{fig:mohr} Example of the Mohr's circle plot of the triaxial tests without wire. The line fits the data with a slope of 16$^\circ$.} }
\end{figure}

\begin{figure}[htb]
\centering{
\includegraphics[width=8cm]{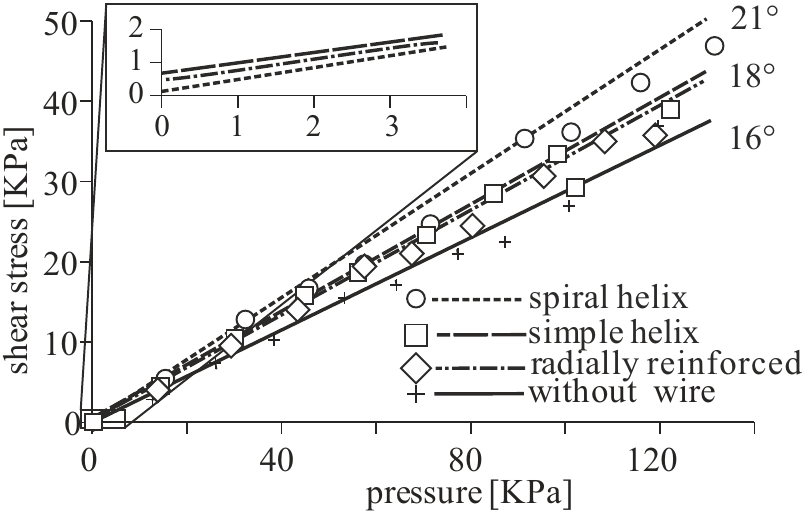}
\caption{\label{fig:angle}Comparison of the Mohr-Coulomb yield surfaces with friction angles for each pattern. The inset shows the intersection with the vertical axis.} }
\end{figure}

The resulting Mohr plot for the spiral helix configuration is shown in Fig.~\ref{fig:mohr}. The slope of the fit through shear stress maxima gives the friction angle while the intercept with the vertical axis gives the cohesion, hence the strength of the material in the absence of any confining pressure. In Fig.~\ref{fig:angle}, we compare the performance of the different patterns. Without any wire, the material exhibits the lowest angle of friction with 16$^\circ$ and no cohesion as it is only composed of spherical particles. All configurations with wire have significantly higher friction angles. From our experiments we find that radial reinforcements do not increase the strength of the material much, compared to the simple helix pattern. The spiral helix pattern performs the best of all studied patterns with a friction angle of 21$^\circ$ and a cohesion of 1~KPa.

\subsection{Energy distribution in the wire}
\begin{figure*}[htb]
\centering{
\includegraphics[width=16.5cm]{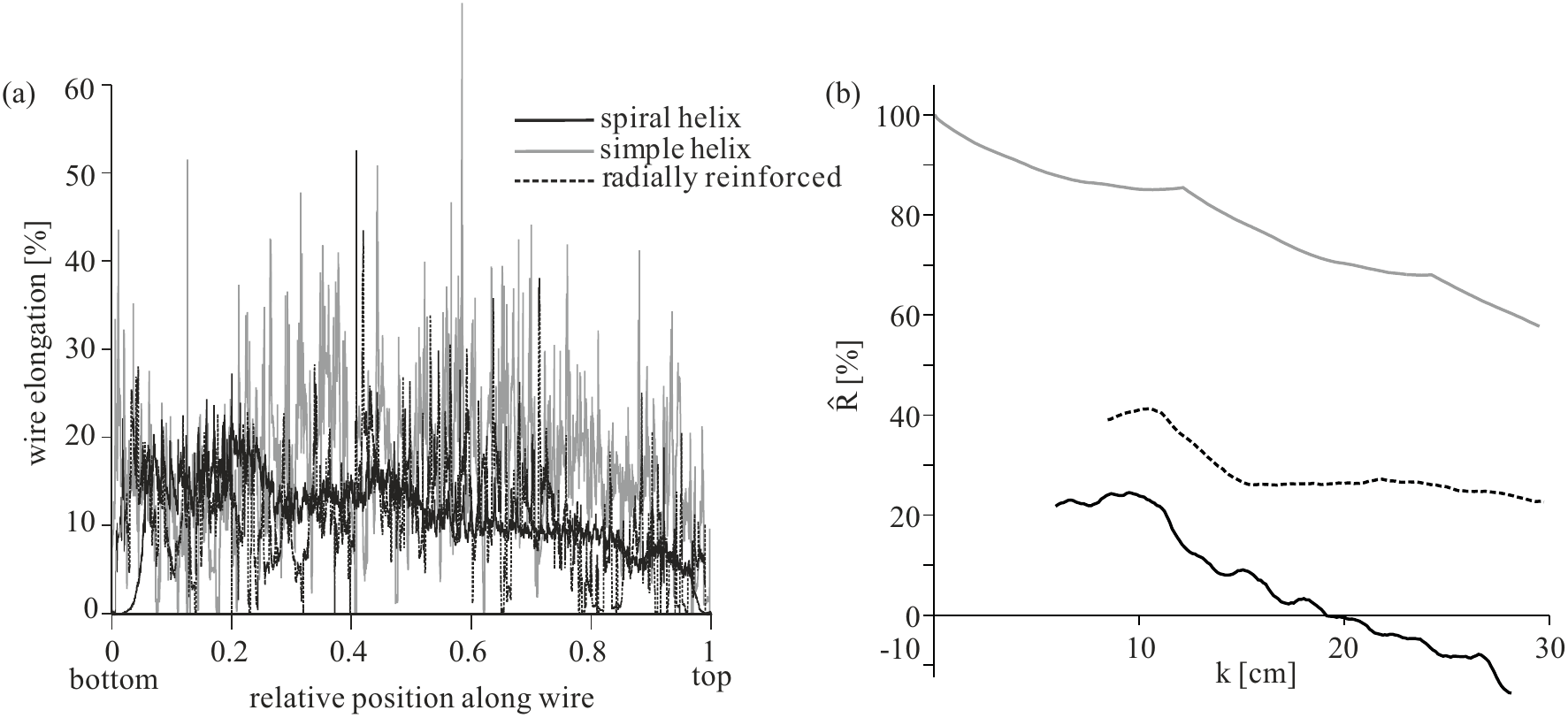}
\caption{\label{fig:autocorrelation} (a) Elongation of the wire for the different configurations and (b) autocorrelation of the elongation for the different configurations with identical legend.} }
\end{figure*}
To obtain a deeper insight into role of the wire inside the granular packing, we plot in Fig.~\ref{fig:autocorrelation}(a) the elongation of wire segments along the total length of the deployed wire for the different configurations. Note that the end points are not anchored. For the best performing configuration, i.e. the spiral helix, the high peaks show that a few loops have strains up to 40\%, what is about twice the average value. It is also interesting to observe periodic drops to nearly 10\% between the loops. This suggests that a fully continuous wire reinforcement of the granular structure might not be necessary for the spiral helix, as cuts at those locations would probably not strongly influence the performance. In Fig.~\ref{fig:autocorrelation}(b), we plot the normalized autocorrelation coefficients of the elongation, i.e. the correlation of the signal $X$ with itself shifted by an offset $k$, called lag, on the horizontal axis. We define the autocorrelation function as
\begin{equation}\label{eq:autocorrelation}
 \hat{R}(k)=\frac{1}{(N^w-k) \sigma^2_{elo}} \sum_{t=1}^{N^w-k} (X_t-\mu_{elo})(X_{t+k}-\mu_{elo}),
\end{equation}
with $N^w$ wire segments, the variance $\sigma^2_{elo}$, and the mean wire segment elongation $\mu_{elo}$ . The maximum autocorrelation is found for a periodicity of 9.5~cm (cf. Fig.~\ref{fig:autocorrelation}(b)) for the spiral helix, what is slightly less than the 10~cm perimeter of the deposited loops. A similar periodicity of the wire is not found for the simple helix configuration and is less marked for the radial reinforced helix pattern.
\section{Conclusions and Outlook}
It can be concluded even from the limited number of simulations, that the deployment pattern has a crucial influence on the performance of continuous wire reinforced granular systems. The robustness of the different patterns with respect to wire failures or slipping of strongly elongated wire segments is quite different for the different deployment strategies, with the simple helix being the least robust one. As the particles are poured, interactions with the wire deposition may prevent the packing in some regions to reach a density high enough for a robust configuration under load, in particular for high wire volume fractions. Note that this effect is mainly due to a choice of a rather large wire radius with respect to the particle one. For high volume fractions, the ability of the granular system to rearrange force chains is hindered, changing the qualitative behavior of the systems upon failure, for example from a ductile collapse to a solid fracture. The comparison of the spiral helix with the radially reinforced configuration shows the disadvantage of deployment patterns with trajectories containing regions with large curvatures.

These observations point at the fact that there might be an optimum reinforcement strategy with respect to particle size and deployment parameters that needs to be explored in the future by an extensive parameter study. These should involve not only parameters of triaxial tests, but also robustness of loaded configurations with respect to wire failure, size effects and wire properties. Novel measurement methods based on string sensors can give insight into the load distribution along the wire in real systems. Spheres resemble a strong idealization with respect to particle shapes that will be overcome in follow-up works by incorporating polyhedral particles with advanced contact calculation schemes \cite{X1,X2,X3,X4}.
\section*{Acknowledgements}
Financial support from ETH Zurich by ETHIIRA Grant No. ETH-04 14-2 as well as from the European Research Council (ERC) Advanced Grant No. 319968-FlowCCS is gratefully acknowledged.


\end{document}